\documentclass[11pt]{article}%
\usepackage{geometry}
\usepackage{dsfont}
\usepackage{amsmath}
\usepackage{amsfonts}
\usepackage{amssymb}
\usepackage{graphicx}%
\usepackage{subfig}
\usepackage{float,pslatex}

\newcommand{\p}{\partial}

\newcommand{\F}{\vphantom{x}_2F_1}
\newcommand{\G}{\vphantom{x}_3F_2}

\renewcommand{\d}{\delta}
\renewcommand{\S}{\Sigma}

\renewcommand{\d}{{\mathrm{d}}}

\renewcommand{\d}{{\mathrm{d}}}

\renewcommand{\p}{\partial}

\renewcommand{\F}{\ensuremath{\mathcal{F}}}
\newcommand{\h}{\ensuremath{\mathcal{H}}}
\renewcommand{\G}{\ensuremath{\mathcal{G}}}

\newcommand{\GZ}{\ensuremath{\mathrm{GZ}}}
\newcommand{\oG}{\ensuremath{\overline{\mathcal{G}}}}
\newcommand{\oF}{\ensuremath{\overline{\mathcal{F}}}}

\begin{document}

\date{}
\title{\textbf{Spontaneous breaking of the BRST symmetry in presence of the Gribov horizon: renormalizability}}
\author{\textbf{M.~A.~L.~Capri}$^{a}$\thanks{caprimarcio@gmail.com}\,\,,
\textbf{D.~Dudal}$^{b}$\thanks{david.dudal@ugent.be}\,\,,
\textbf{M.~S.~Guimaraes}$^{a}$\thanks{msguimaraes@uerj.br}\,\,,
\textbf{I.~F.~Justo}$^{a}$\thanks{igorfjusto@gmail.com}\,\,,
\textbf{L.~F.~Palhares }$^{a,c}$\thanks{palhares@thphys.uni-heidelberg.de}\,\,,\\
\textbf{S.~P.~Sorella}$^{a}$\thanks{sorella@uerj.br}\ \thanks{Work supported by
FAPERJ, Funda{\c{c}}{\~{a}}o de Amparo {\`{a}} Pesquisa do Estado do Rio de
Janeiro, under the program \textit{Cientista do Nosso Estado}, E-26/101.578/2010.}\,\,\\[2mm]
{\small \textnormal{$^{a}$  \it Departamento de F\'{\i }sica Te\'{o}rica, Instituto de F\'{\i }sica, UERJ - Universidade do Estado do Rio de Janeiro,}}
 \\ \small \textnormal{\phantom{$^{a}$} \it Rua S\~{a}o Francisco Xavier 524, 20550-013 Maracan\~{a}, Rio de Janeiro, Brasil}\\
	 \small \textnormal{$^{b}$ \it Ghent University, Department of Physics and Astronomy, Krijgslaan 281-S9, 9000 Gent, Belgium}\\
	 \small \textnormal{$^{c}$ \it Institut f\"ur Theoretische Physik, Heidelberg University, Philosophenweg 16,  69120 Heidelberg, Germany}\normalsize}
\maketitle
\begin{abstract}
An all orders algebraic proof of the multiplicative renormalizability of the novel formulation of the Gribov-Zwanziger action proposed in \cite{Dudal:2012sb}, and allowing for an exact  but  spontaneously broken BRST symmetry, is provided.
\end{abstract}

\baselineskip=13pt



\maketitle
\section{Introduction}
In recent years much attention has been devoted to the study of the issue of the Gribov copies  \cite{Gribov:1977wm} and of its relevance for confinement in Yang-Mills theories\footnote{See refs.\cite{Sobreiro:2005ec,Vandersickel:2012tz} for a pedagogical introduction to the Gribov problem.} . The existence of the Gribov copies is a general feature of the gauge fixing quantization procedure, being related to the impossibility of finding a local gauge condition which picks up only one gauge configuration for each gauge orbit  \cite{Singer:1978dk}. As it has been shown by  Gribov and Zwanziger \cite{Gribov:1977wm,Zwanziger:1989mf,Zwanziger:1992qr}, a partial resolution of the Gribov problem in the Landau gauge can be achieved by  restricting the domain of integration in the functional Euclidean integral to the first Gribov horizon. Remarkably, this restriction has resulted into  a local and renormalizable action,  known as the Gribov-Zwanziger action \cite{Zwanziger:1989mf,Zwanziger:1992qr}.

\noindent More recently,  a  Refined version of the Gribov-Zwanziger action has been worked out in \cite{Dudal:2007cw,Dudal:2008sp,Dudal:2011gd}, leading to a tree level gluon propagator whose behavior in the infrared region is in very good agreement with the most recent lattice numerical simulations \cite{Cucchieri:2007rg,Bornyakov:2008yx,Dudal:2010tf,Cucchieri:2011ig,Oliveira:2012eh,Dudal:2012zx}. This propagator displays  complex poles in momentum space. As such, it cannot describe the propagation of physical excitations. Rather, it is suited for  a kind of effective description of  gluon confinement, see also \cite{Stingl:1994nk}.  In spite of the appearance of complex poles, the Refined-Gribov-Zwanziger gluon propagator has been successfully employed to investigate the correlation functions of gauge invariant composite glueball operators in order to get estimates of the glueball masses. It turns out that a K{\"a}ll{\'e}n-Lehmann spectral representation with positive spectral density can be extracted from these correlation functions \cite{Baulieu:2009ha},with numerical verification recently considered in \cite{Windisch:2012sz}. The resulting mass estimates for the lowest glueball states with quantum numbers $J^{PC}=0^{++}, 2^{++}, 0^{-+}$, are in qualitative agreement with the available numerical data on the spectrum of the glueballs \cite{Dudal:2010cd}. Let us also  mention that such type of gluon propagator has also been used in previous studies in hadron physics \cite{Roberts:1994dr,Bhagwat:2002tx}, see also  \cite{Baulieu:2009xr,Dudal:2013vha} for a recent attempt to generalize the Refined-Gribov-Zwanziger action by including quarks and associated chiral symmetry breaking. Recently, complex pole propagators were also considered in terms of semi-analytical approaches to the QCD phase diagram \cite{Benic:2012ec,Fukushima:2012qa}, partially motivated by fits to finite temperature lattice gluon data \cite{Aouane:2011fv,Cucchieri:2011di}.

\noindent Although the aforementioned  results can be taken as evidence of the fact that the Refined Gribov-Zwanziger theory can be effectively employed to investigate the physical spectrum of a confining Yang-Mills theory, there are still many aspects of the theory which remain to be understood. Certainly, the systematic construction of a set of composite operators whose correlation functions can be directly related to the physical spectrum of a confining Yang-Mills theory is one of the most challenging aspects of the Gribov-Zwanziger framework for color confinement. At present, the characterization  of the analyticity and of the unitarity properties of these correlation functions seems a highly cumbersome task,  taking into account that explicit calculations have to be done by employing a confining gluon propagator exhibiting complex poles.

\noindent Amongst the various open aspects of the Gribov-Zwanziger framework, the issue of the BRST symmetry is a source of continuous investigations, see for example \cite{Zwanziger:1989mf,Zwanziger:1992qr,Dudal:2008sp,Maggiore:1993wq,Baulieu:2008fy,Sorella:2009vt,Dudal:2010hj,Dudal:2009xh,Capri:2010hb,Capri:2011wp,vonSmekal:2007ns,vonSmekal:2008en,Serreau:2012cg, Lavrov:2011wb} for an overview of what has been already done on this topic. We expect that a better understanding of the role of the BRST symmetry in confining Yang-Mills theories would be of great relevance in order to face the characterization of the physical spectrum.

\noindent In a recent work \cite{Dudal:2012sb}, some of the authors have been able to obtain an equivalent formulation of the Gribov-Zwanziger action which displays an exact BRST symmetry which turns out to be spontaneously broken by the restriction of the domain of integration to the Gribov horizon. In particular, in \cite{Dudal:2012sb}, the BRST operator $s$ retains the important property of being nilpotent, {\it i.e.} $s^2=0$. This feature enabled us to make use of the powerful tool of the cohomology of $s$ \cite{Barnich:2000zw,Piguet:1995er} in order to prove that the set of colorless gauge invariant operators corresponding to the cohomology classes of $s$ is closed under time evolution \cite{Dudal:2012sb}. Moreover, it has also been shown that the Goldstone mode associated to the spontaneous breaking of $s$ is completely decoupled.

\noindent The aim of the present article is to fill a gap not addressed in the previous work  \cite{Dudal:2012sb}, namely,  the renormalizability to all orders of the spontaneous symmetry breaking formulation of the Gribov-Zwanziger theory in $d=4$. As we shall see, the action obtained in \cite{Dudal:2012sb} enjoys a large set of Ward identities which enables us to prove that it is, in fact,  multiplicatively renormalizable to all orders.

\noindent The paper is organized as follows. In Sect.~2 we provide a short summary of the BRST spontaneous symmetry breaking formulation of the Gribov-Zwanziger action. In Sect.~3 we derive numerous Ward identities fulfilled by the action in the novel formulation. In Sect.~4,  the renormalizability to all orders  of the model is established by means of the algebraic renormalization procedure \cite{Piguet:1995er}.

\section{A novel formulation of the Gribov-Zwanziger action and the spontaneous breaking of the BRST symmetry}
Let us start by recalling the expression of the Gribov-Zwanziger action which enables us to restrict the Euclidean functional integral to the first Gribov horizon, namely
\begin{eqnarray}\label{GZaction}
S_{GZ} &=& \frac{1}{4}\int d^4x\; F^a_{\mu\nu} F^a_{\mu\nu}+\int d^4 x\,\left( i b^a \p_\mu A_\mu^a +\overline c^a \p_\mu D_\mu^{ab} c^b \right)  \nonumber \\
&+&\int d^4 x \left( \overline{\varphi }_{\mu }^{ac} \partial _{\nu} D_\nu^{am}\varphi _{\mu
}^{mc} -\overline{\omega }_{\mu }^{ac} \partial _{\nu } D_\nu^{am} \omega _{\mu }^{mc}
  -g\left( \partial _{\nu }\overline{\omega }_{\mu}^{ac}\right) f^{abm}\left( D_{\nu }c\right) ^{b}\varphi _{\mu
}^{mc}\right)\nonumber\\
&+&\int d^4 x\left( -\gamma ^{2}gf^{abc}A_{\mu }^{a}(\varphi _{\mu }^{bc}+\overline{\varphi }_{\mu }^{bc}) - 4\left(N^{2}-1\right) \gamma^{4} \right) \;.
\end{eqnarray}
The field $b^a$ stands for the Lagrange multiplier implementing the Landau gauge condition, $ \p_\mu A_\mu^a = 0$, and $\overline c^a, c^a$ are the corresponding Faddeev-Popov ghosts. The fields  $\overline{\varphi }_{\mu }^{ab}, {\varphi }_{\mu }^{ab}$ are a pair of bosonic fields, while $\overline{\omega }_{\mu }^{ab}, {\omega }_{\mu }^{ab}$ are a pair of anticommuting fields. These fields are needed in order to implement the restriction to the first Gribov horizon in a local way \cite{Zwanziger:1989mf,Zwanziger:1992qr,Dudal:2008sp}. All fields belong to the adjoint representation of the gauge group $SU(N)$. The massive parameter $\gamma^2$, called the Gribov parameter,  is not a free parameter, being determined in a self-consistent way through the gap equation
\begin{equation}
\frac{\p E_{vac}}{\p \gamma^2}  =0 \;, \label{geqs}
\end{equation}
where $E_{vac}$ stands for the vacuum energy of the theory \cite{Gribov:1977wm,Zwanziger:1989mf,Zwanziger:1992qr},
\begin{equation}
e^{-E_{vac}} = \int [D\Phi] \; e^{-S_{GZ}} \;. \label{ev}
\end{equation}
It turns out that the Gribov-Zwanziger action \eqref{GZaction} does not exhibits an exact BRST invariance \cite{Zwanziger:1989mf,Zwanziger:1992qr,Dudal:2008sp}, which is softly broken by the Gribov parameter $\gamma^2$.  Indeed,
\begin{eqnarray}
sS_{GZ}&=& -\gamma^2\int d^4 x\left(gf^{abc}D_\mu^{ak}c^k(\overline\varphi_\mu^{bc}+\varphi_\mu^{bc})+gf^{abc}A_\mu^a\omega_\mu^{bc}\right)  \;,
\end{eqnarray}
where $s$ stands for the nilpotent BRST operator
\begin{eqnarray}
sA_{\mu }^{a}& =&-D_{\mu }^{ab}c^{b}=-(\partial _{\mu }\delta
^{ab}+gf^{acb}A_{\mu }^{c})c^{b}\;,\qquad sc^{a}~=~\frac{g}{2}f^{acb}c^{b}c^{c}\;,\qquad s\overline{c}^{a}~=~i b^{a}\;,\qquad sb^{a}~=~0\;, \nonumber \\
s\overline{\omega}_{\mu }^{ab}& =&\overline{\varphi}_{\mu }^{ab}\;,\qquad s\overline{\varphi}_{\mu }^{ab}~=~0\;,\qquad s\varphi _{\mu }^{ab}~=~\omega _{\mu }^{ab}\;,\qquad s\omega _{\mu
}^{ab}~=~0\;,  \qquad s^2~=~0 \;. \label{BRS}
\end{eqnarray}
In  ref.\cite{Dudal:2012sb} it was proposed to replace expression \eqref{GZaction} by the following action
\begin{eqnarray}\label{GZaction2}
S'_{GZ} &=& \frac{1}{4}\int d^4 x\; F^a_{\mu\nu} F^a_{\mu\nu}+\int d^4 x\,\left(i b^a \p_\mu A_\mu^a +\overline c^a \p_\mu D_\mu^{ab} c^b \right)\nonumber\\
&&+\int d^4 x \left( \overline{\varphi }_{\mu }^{ac} \partial _{\nu} D_\nu^{am}\varphi _{\mu
}^{mc} -\overline{\omega }_{\mu }^{ac} \partial _{\nu } D_\nu^{am} \omega _{\mu }^{mc}
  -g\left( \partial _{\nu }\overline{\omega }_{\mu}^{ac}\right) f^{abm}\left( D_{\nu }c\right) ^{b}\varphi _{\mu
}^{mc}\right)\nonumber\\
&&+\int d^4 x\left(- \oG_{\mu\nu}^{ab}\p^2\G_{\mu\nu}^{ab}+ \oF_{\mu\nu}^{ab}\p^2\F_{\mu\nu}^{ab}- \oG_{\mu\nu}^{ab}D_{\mu}^{ak}\overline\varphi_\nu^{kb}-gf^{ak\ell }\oG_{\mu\nu}^{ab}D_\mu^{\ell p}c^p\overline\omega_\nu^{k b}  \right) \nonumber\\
&&+\int d^4 x\left(- \widehat{\oG}_{\mu\nu}^{ab}\p^2\widehat{\G}_{\mu\nu}^{ab}+ \widehat{\oF}_{\mu\nu}^{ab}\p^2\widehat{\F}_{\mu\nu}^{ab}+\widehat{\oF}_{\mu\nu}^{ab}D_\mu^{ak}\omega^{kb}_\nu- \widehat{\oG}_{\mu\nu}^{ab}D_{\mu}^{ak}\varphi_\nu^{kb}+gf^{a k\ell}\widehat{\oF}_{\mu\nu}^{ab}D_\mu^{\ell p}c^p\varphi_\nu^{k b} \right)\nonumber\\&&
+\int d^4 x\left(\h_{\mu\nu}^{ab}\left(\oG_{\mu\nu}^{ab}-\delta_{\mu\nu}\delta^{ab}\gamma^2\right)+\widehat{\h}_{\mu\nu}^{ab}\left(\widehat{\oG}_{\mu\nu}^{ab}-\delta_{\mu\nu}\delta^{ab}\gamma^2\right)-\oG_{\mu\nu}^{ab}\widehat{\oG}_{\mu\nu}^{ab}\right)  \;,
\end{eqnarray}
where we have  introduced two new BRST quartets   \cite{Piguet:1995er} of fields consisting of $\F_{\mu\nu}^{ab}$, $\oF_{\mu\nu}^{ab}$ (commuting), $\G_{\mu\nu}^{ab}$, $\oG_{\mu\nu}^{ab}$ (anticommuting) and their hat-counterparts:
\begin{eqnarray}\label{quartet1}
    s\overline \F_{\mu\nu}^{ab}&=&\oG_{\mu\nu}^{ab}\,,\qquad s \oG_{\mu\nu}^{ab} ~=~ 0\,,\qquad s \widehat{\oF}_{\mu\nu}^{ab}=\widehat{\oG}_{\mu\nu}^{ab}\,,\qquad s \widehat{\oG}_{\mu\nu}^{ab} ~=~ 0\,,\nonumber\\
    s \G_{\mu\nu}^{ab}&=&\F_{\mu\nu}^{ab}\,,\qquad s \F_{\mu\nu}^{ab}~=~0\,,\qquad  s\widehat{\G}_{\mu\nu}^{ab}=\widehat{\F}_{\mu\nu}^{ab}\,,\qquad s \widehat{\F}_{\mu\nu}^{ab}~=~0  \;, \label{BRS1}
\end{eqnarray}
as well as the singlet fields  $\h_{\mu\nu}^{ab},\widehat{\h}_{\mu\nu}^{ab}$
\begin{equation}
s\h_{\mu\nu}^{ab}=s\widehat{\h}_{\mu\nu}^{ab}=0 \;.  \label{singlets}
\end{equation}
It is easily checked that the action \eqref{GZaction2} can be  rewritten as
\begin{eqnarray}\label{GZaction2bis}
S'_\GZ &=& \frac{1}{4}\int d^4 x\; F^a_{\mu\nu} F^a_{\mu\nu}+ s \int d^4 x\,\left( \overline c^a \p_\mu A_\mu^a +\overline{\omega }_{\mu }^{ac} \partial _{\nu} D_\nu^{am}\varphi _{\mu
}^{mc}\right)\nonumber\\
&&+ s \int d^4 x\,\left(- \oF_{\mu\nu}^{ab}\p^2\G_{\mu\nu}^{ab}- \oG_{\mu\nu}^{ab}D_{\mu}^{ak}\overline\omega_\nu^{kb}  - \widehat{\oF}_{\mu\nu}^{ab}\p^2\widehat{\G}_{\mu\nu}^{ab}- \widehat{\oF}_{\mu\nu}^{ab}D_{\mu}^{ak}\varphi_\nu^{kb} \right)\nonumber\\&&
+\int d^4 x\left(\h_{\mu\nu}^{ab}\left(\oG_{\mu\nu}^{ab}-\delta_{\mu\nu}\delta^{ab}\gamma^2\right)+\widehat{\h}_{\mu\nu}^{ab}\left(\widehat{\oG}_{\mu\nu}^{ab}-\delta_{\mu\nu}\delta^{ab}\gamma^2\right)-\oG_{\mu\nu}^{ab}\widehat{\oG}_{\mu\nu}^{ab}\right)\;,
\end{eqnarray}
from which it can be established that $S'_\GZ$ has an exact BRST invariance
\begin{equation}
sS'_{GZ}=0   \;, \label{excat}
\end{equation}
whereby we have preserved the nilpotency of the BRST operator, $s^2=0$.

\noindent As discussed  in  \cite{Dudal:2012sb}, the formulation provided by the novel action $S'_\GZ$ is equivalent to that of the original Gribov-Zwanziger action $S_{GZ}$. Let us therefore point out that, using the algebraic exact equations of motion of the fields $(\h,\widehat{\h})$,
\begin{equation}
\oG_{\mu\nu}^{ab}=\widehat{\oG}_{\mu\nu}^{ab}=\gamma^2\delta^{ab}\delta_{\mu\nu} \;, \label{heom}
\end{equation}
we immediately recover the $\gamma$-dependent part of the Gribov-Zwanziger action, namely
\begin{eqnarray}\label{GZaction2tris}
&&\int d^4 x\left(- \oG_{\mu\nu}^{ab}D_{\mu}^{ak}\overline\varphi_\nu^{kb}- \widehat{\oG}_{\mu\nu}^{ab}D_{\mu}^{ak}\varphi_\nu^{kb}- \oG_{\mu\nu}^{ab}\p^2\G_{\mu\nu}^{ab}- \widehat{\oG}_{\mu\nu}^{ab}\p^2\widehat{\G}_{\mu\nu}^{ab}-\oG_{\mu\nu}^{ab}\widehat{\oG}_{\mu\nu}^{ab}+\h_{\mu\nu}^{ab}\left(\oG_{\mu\nu}^{ab}-\delta_{\mu\nu}\delta^{ab}\gamma^2\right)\right)\nonumber\\
&&\stackrel{\h-,\widehat{\h}-EOM}{\longrightarrow}
\int d^4\left(- \gamma^2 gf^{abc}A_\mu^a \left(\varphi_\mu^{bc}+\overline\varphi_\mu^{bc}\right)-4(N^2-1)\gamma^4\right)\,.
\end{eqnarray}
Moreover, the integration over the fields $\F$, $\oF$, $\widehat{\F}$ and $\widehat{\oF}$ turns out to generate a unity in the partition function, as can be  seen by performing the following  simultaneous changes  of integration variables:
\begin{eqnarray}
\widehat{\F}_{\mu\nu}^{ab}&\to&\widehat{\F}_{\mu\nu}^{ab}-\frac{1}{\p^2}\left(D_\mu^{ak}\omega_{\nu}^{kb}+gf^{a k\ell}D_\mu^{\ell p}c^p\varphi_\nu^{k b}\right)\,,\qquad
\omega_{\nu}^{tb}~\to~\omega_{\nu}^{tb}+\left[(\p D)^{-1}\right]^{tk}\left(gf^{ak\ell} \oG_{\mu\nu}^{ab}D_\mu^{\ell p}c^p\right)\,,\nonumber\\
\F_{\mu\nu}^{ab}&\to&\F_{\mu\nu}^{ab}-\frac{1}{\p^2}\left(D_\mu^{ak}\omega_\nu^{kb}+gf^{a k\ell}D_\mu^{\ell p}c^p\overline\varphi_\nu^{k b}\right) \;,
\end{eqnarray}
from which the equivalence between the two formulations, expressed by means of \eqref{GZaction} and \eqref{GZaction2}, follows.

\noindent Let us proceed by showing how the spontaneous breaking of the BRST symmetry is realized in the  new formulation of the Gribov-Zwanziger action. To that end, let us rewrite the action $S'_\GZ$  by again making explicit use of the equations of motion \eqref{heom}. Thus
\begin{eqnarray}\label{GZactioneom}
S'_\GZ &=& \frac{1}{4}\int d^4 x\; F^a_{\mu\nu} F^a_{\mu\nu}+\int d^4 x\,s\left( \overline c^a \p_\mu A_\mu^a +\overline{\omega }_{\mu }^{ac} \partial _{\nu} D_\nu^{am}\varphi _{\mu
}^{mc}\right)\nonumber\\
&&+\int d^4 x\;  \left( \oF_{\mu\nu}^{ab}\p^2\F_{\mu\nu}^{ab}-   \gamma^2 \; s (D_{\mu}^{ak}\overline\omega_\mu^{ka})
 + \widehat{\oF}_{\mu\nu}^{ab}\p^2\widehat{\F}_{\mu\nu}^{ab}- \gamma^2 D_{\mu}^{ak}\varphi_\mu^{ka}
 +  \widehat{\oF}_{\mu\nu}^{ab} \;s (D_{\mu}^{ak}\varphi_\nu^{kb})  - \gamma^4 d (N^2-1)  \right) \;. \nonumber \\
\end{eqnarray}
This expression  turns out  to be left invariant by the following nilpotent BRST transformations:
\begin{eqnarray}
sA_{\mu }^{a}& =&-D_{\mu }^{ab}c^{b}=-(\partial _{\mu }\delta
^{ab}+gf^{acb}A_{\mu }^{c})c^{b}\;,  \qquad sc^{a}~=~\frac{g}{2}f^{acb}c^{b}c^{c}\;,\qquad s\overline{c}^{a}~=~i b^{a}\;,\qquad sb^{a}~=~0\;,  \nonumber \\
s\overline{\omega}_{\mu }^{ab}& =&\overline{\varphi}_{\mu }^{ab}\;,\qquad s\overline{\varphi}%
_{\mu }^{ab}~=~0\;,  \qquad s\varphi _{\mu }^{ab}~=~\omega _{\mu }^{ab}\;,\qquad s\omega _{\mu
}^{ab}~=~0\;,   \nonumber\\
s\oF_{\mu\nu}^{ab} & =& \gamma^2 \delta^{ab}\delta_{\mu\nu}\;, \qquad  s  \widehat{\oF}_{\mu\nu}^{ab}~=~ \gamma^2 \delta^{ab}\delta_{\mu\nu} \;, \qquad s \G_{\mu\nu}^{ab}~=~ \F_{\mu\nu}^{ab}\;, \qquad s \F_{\mu\nu}^{ab}~=~ 0\;, \qquad \nonumber \\
s\widehat{\G}_{\mu\nu}^{ab}~& = & ~\widehat{\F}_{\mu\nu}^{ab}\,,\qquad s \widehat{\F}_{\mu\nu}^{ab}~=~0  \;, \label{BRSTssb}
\end{eqnarray}
with
\begin{equation}
s S'_\GZ = 0\,,\qquad s^2  =0 \;. \label{ssbinv}
\end{equation}
Furthermore, from equations \eqref{BRSTssb}, it follows that the BRST operator suffers from spontaneous symmetry breaking. In fact
\begin{equation}
\langle s\oF_{\mu\nu}^{ab} \rangle  = \gamma^2 \delta^{ab}\delta_{\mu\nu} \;, \qquad  \langle s  \widehat{\oF}_{\mu\nu}^{ab} \rangle = \gamma^2 \delta^{ab}\delta_{\mu\nu} \;.  \label{s_ssb}
\end{equation}
Let us end this short summary by mentioning the important feature that the Goldstone mode associated to the spontaneous symmetry breaking of the BRST operator turns out to be completely decoupled from the theory, see  \cite{Dudal:2012sb} for the argument.

\section{Ward identities}
The first step in order to prove the all orders renormalizability of the novel formulation is to establish the set of Ward identities  obeyed by the action $S'_{GZ}$, eq.\eqref{GZaction2}. To that end, and following the algebraic renomalization procedure \cite{Piguet:1995er}, we introduce a set of external sources $\lambda^{ab}_{i}, \rho^{ab}_{i}, K^a_\mu, L^a$ transforming as
\begin{equation}
s\lambda^{ab}_{i}=\rho^{ab}_{i}\,,\qquad s\rho^{ab}_{i}=0\;, \qquad sK^{a}_{\mu}= s L^a = 0  \;,  \label{sources}
\end{equation}
and the complete BRST invariant action $\Sigma$
\begin{eqnarray}
\Sigma&=&\int d^{4}x\,\bigg\{\frac{1}{4}F^{a}_{\mu\nu}F^{a}_{\mu\nu}
+ib^{a}\,\partial_{\mu}A^{a}_{\mu}
+\bar{c}^{a}\,\partial_{\mu}D^{ab}_{\mu}c^{b}
+\bar{\phi}^{a}_{i}\,\partial_{\mu}D^{ab}_{\mu}\phi^{b}_{i}
-\bar{\omega}^{a}_{i}\,\partial_{\mu}D^{ab}_{\mu}\omega^{b}_{i}
-gf^{abc}(\partial_{\mu}\bar\omega^{a}_{i})(D^{bd}_{\mu}c^{d})\phi^{b}_{i}\nonumber\\
&&-{\overline{\mathcal{G}}}^{a}_{\mu i}\,\partial^{2}\mathcal{G}^{a}_{\mu i}
+{\overline{\mathcal{F}}}^{a}_{\mu i}\,\partial^{2}\mathcal{F}^{a}_{\mu i}
-{\overline{\mathcal{G}}}^{a}_{\mu i}\,D^{ab}_{\mu}\bar{\phi}^{b}_{i}
+gf^{abc}{\overline{\mathcal{G}}}^{a}_{\mu i}(D^{bd}_{\mu}c^{d})\bar{\omega}^{c}_{i}
-{\widehat{\overline{\mathcal{G}}}}^{a}_{\mu i}\,\partial^{2}{\widehat{\mathcal{G}}}^{a}_{\mu i}
+{\widehat{\overline{\mathcal{F}}}}^{a}_{\mu i}\,\partial^{2}{\widehat{\mathcal{F}}}^{a}_{\mu i}
+{\widehat{\overline{\mathcal{F}}}}^{a}_{\mu i}\,D^{ab}_{\mu}{\omega}^{b}_{i}\nonumber\\
&&-{\widehat{\overline{\mathcal{G}}}}^{a}_{\mu i}\,D^{ab}_{\mu}{\phi}^{b}_{i}
-gf^{abc}{\widehat{\overline{\mathcal{F}}}}^{a}_{\mu i}(D^{bd}_{\mu}c^{d}){\phi}^{c}_{i}
+\mathcal{H}^{a}_{\mu i}\left({\overline{\mathcal{G}}}^{a}_{\mu i}-\delta^{a}_{\mu i}\gamma^{2}\right)
+{\widehat{\mathcal{H}}}^{a}_{\mu i}\left({\widehat{\overline{\mathcal{G}}}}^{a}_{\mu i}-\delta^{a}_{\mu i}\gamma^{2}\right)
-{\overline{\mathcal{G}}}^{a}_{\mu i}{\widehat{\overline{\mathcal{G}}}}^{a}_{\mu i}\nonumber\\
&&-K^{a}_{\mu}\,D^{ab}_{\mu}c^{b}
+\frac{g}{2}f^{abc}L^{a}c^{b}c^{c}
+\rho^{ab}_{i}\,{\widehat{\overline{\mathcal{F}}}}^{a}_{\mu i}\,D^{bc}_{\mu}c^{c}
-\lambda^{ab}_{i}\,{\widehat{\overline{\mathcal{G}}}}^{a}_{\mu i}\,D^{bc}_{\mu}c^{c}\biggr\}\,,   \label{acc}
\end{eqnarray}
\begin{equation}
 s \Sigma = 0 \;, \label{sS}
\end{equation}
where, as done in the original work by Zwanziger \cite{Zwanziger:1989mf,Zwanziger:1992qr}, we have introduced the multi-index notation $i\equiv(a,\mu)$, $i=1,...,f=4(N^2-1)$, which turns out to be very useful in the discussion of the renormalizability. As pointed out in \cite{Zwanziger:1989mf,Zwanziger:1992qr}, the possibility of introducing the multi-index $i\equiv(a,\mu)$ relies on the existence of a global symmetry $U(f)$.  Thus, the term $\delta^{a}_{\mu i}$ appearing in expression \eqref{acc} stands for
\begin{equation}
\delta^{a}_{\mu i}\equiv\delta^{ab}\delta_{\mu\nu}\,.
\end{equation}
As one can see from the expression $\Sigma$, the external sources $K^a_\mu,L^a$ are introduced in order to properly define the composite operators $D_{\mu }^{ab}c^{b}$  and $\frac{g}{2}f^{acb}c^{b}c^{c}$, corresponding to the nonlinear BRST transformations of the fields $A^a_\mu$ and $c^a$, eqs.\eqref{BRS}. Moreover, it turns out to be useful to also couple the BRST doublet of external fields $\lambda^{ab}_{i}, \rho^{ab}_{i}$ to the composite operators ${\widehat{\overline{\mathcal{G}}}}^{a}_{\mu i}\,D^{bc}_{\mu}c^{c}$ and ${\widehat{\overline{\mathcal{F}}}}^{a}_{\mu i}\,D^{bc}_{\mu}c^{c}$.

\noindent We are now ready to derive an extensive set of Ward identities fulfilled by the action $\Sigma$. These are:

\vspace{3mm}

$\bullet$ the Slavnov-Taylor identity
\begin{eqnarray}
\mathcal{S}(\Sigma)&\equiv&\int d^{4}x\,\biggl\{
\frac{\delta\Sigma}{\delta{K}^{a}_{\mu}}\frac{\delta\Sigma}{\delta{A}^{a}_{\mu}}
+\frac{\delta\Sigma}{\delta{L}^{a}}\frac{\delta\Sigma}{\delta{c}^{a}}
+ib^{a}\frac{\delta\Sigma}{\delta\bar{c}^{a}}
+\bar\phi^{a}_{i}\frac{\delta\Sigma}{\delta\bar{\omega}^{a}_{i}}
+\omega^{a}_{i}\frac{\delta\Sigma}{\delta{\phi}^{a}_{i}}
+\overline{\mathcal{G}}^{a}_{\mu i}\frac{\delta\Sigma}{\delta{\overline{\mathcal{F}}}^{a}_{\mu i}}
+{\mathcal{F}}^{a}_{\mu i}\frac{\delta\Sigma}{\delta{{\mathcal{G}}}^{a}_{\mu i}}\nonumber\\
&&+{\widehat{\overline{\mathcal{G}}}}^{a}_{\mu i}\frac{\delta\Sigma}{\delta{\widehat{\overline{\mathcal{F}}}}^{a}_{\mu i}}
+{\widehat{\mathcal{F}}}^{a}_{\mu i}\frac{\delta\Sigma}{\delta{\widehat{{\mathcal{G}}}}^{a}_{\mu i}}
+\rho^{ab}_{i}\frac{\delta\Sigma}{\delta\lambda^{ab}_{i}}\biggr\}=0\,.  \label{sti}
\end{eqnarray}

\vspace{3mm}

$\bullet$  the linearly broken Ward identity for the Gribov parameter $\gamma^2$
\begin{equation}
\frac{\partial\Sigma}{\partial\gamma^{2}}=-\int d^{4}x\,\left(\mathcal{H}^{aa}_{\mu\mu}+{\widehat{\mathcal{H}}}^{aa}_{\mu\mu}\right)\,.  \label{ge}
\end{equation}
As we shall see in the next section, this identity will be responsible for the nonrenormalizability properties of the Gribov parameter $\gamma^2$. Notice that the left hand side of \eqref{ge} is linear in the quantum fields, {\it i.e.} it is a linear breaking. It is well established that this kind of breaking is not affected by quantum corrections, see  \cite{Piguet:1995er}.

\vspace{3mm}

$\bullet$ the gauge-fixing condition and the anti-ghost equation:
\begin{equation}
\frac{\delta\Sigma}{\delta{b}^{a}}=\partial_{\mu}A^{a}_{\mu}\,,\qquad
\frac{\delta\Sigma}{\delta\bar{c}^{a}}+\partial_{\mu}\frac{\delta\Sigma}{\delta{K}^{a}_{\mu}}=0\,,
\end{equation}

\vspace{3mm}

$\bullet$  the equations of motion of the auxiliary fields:
\begin{equation}
\frac{\delta\Sigma}{\delta{\mathcal{G}}^{a}_{\mu i}}=-\partial^{2}{\overline{\mathcal{G}}}^{a}_{\mu i}\,,\qquad
\frac{\delta\Sigma}{\delta{\widehat{\mathcal{G}}}^{a}_{\mu i}}=-\partial^{2}{\widehat{\overline{\mathcal{G}}}}^{a}_{\mu i}\,,
\end{equation}
\begin{equation}
\frac{\delta\Sigma}{\delta{\mathcal{F}}^{a}_{\mu i}}=-\partial^{2}{\overline{\mathcal{F}}}^{a}_{\mu i}\,,\qquad
\frac{\delta\Sigma}{\delta{{\overline{\mathcal{F}}}}^{a}_{\mu i}}=\partial^{2}{\mathcal{F}}^{a}_{\mu i}\,,\qquad
\frac{\delta\Sigma}{\delta{\widehat{\mathcal{F}}}^{a}_{\mu i}}=-\partial^{2}{\widehat{\overline{\mathcal{F}}}}^{a}_{\mu i}\,,
\end{equation}

\vspace{3mm}

$\bullet$ the equations of motion of the Lagrange multipliers:
\begin{equation}
\frac{\delta\Sigma}{\delta\mathcal{H}^{a}_{\mu i}}={\overline{\mathcal{G}}}^{a}_{\mu i}-\delta^{a}_{\mu i}\gamma^{2}\,,\qquad
\frac{\delta\Sigma}{\delta{\widehat{\mathcal{H}}}^{a}_{\mu i}}={\widehat{\overline{\mathcal{G}}}}^{a}_{\mu i}-\delta^{a}_{\mu i}\gamma^{2}\,,
\end{equation}

\vspace{3mm}

$\bullet$ the equations of motion of the localizing fields:
\begin{eqnarray}
\overline{\Phi}^{a}_{i}(\Sigma)&\equiv&\frac{\delta\Sigma}{\delta\bar\phi^{a}_{i}}
+\partial_{\mu}\frac{\delta\Sigma}{\delta{\widehat{\overline{\mathcal{G}}}}^{a}_{\mu i}}
-D^{ab}_{\mu}\frac{\delta\Sigma}{\delta\mathcal{H}^{b}_{\mu i}}
-\partial_{\mu}\left(\lambda^{ab}_{i}\frac{\delta\Sigma}{\delta{K}^{b}_{\mu}}\right)\nonumber\\
&&=-\partial^{2}\partial_{\mu}{\widehat{\mathcal{G}}}^{a}_{\mu i}
+\partial_{\mu}{\widehat{\mathcal{H}}}^{a}_{\mu i}
+\partial_{\mu}{\overline{\mathcal{G}}}^{a}_{\mu i}
+\gamma^{2}gf^{abc}A^{c}_{\mu}\delta^{b}_{\mu i}\,,\\\cr
\overline{\Omega}^{a}_{i}(\Sigma)&\equiv&
\frac{\delta\Sigma}{\delta\bar\omega^{a}_{i}}
+\partial_{\mu}\frac{\delta\Sigma}{\delta{\widehat{\overline{\mathcal{F}}}}^{a}_{\mu i}}
-gf^{abc}\left(\frac{\delta\Sigma}{\delta\mathcal{H}^{b}_{\mu i}}+\delta^{b}_{\mu i}\gamma^{2}\right)\frac{\delta\Sigma}{\delta{K}^{c}_{\mu}}
+\partial_{\mu}\left(\rho^{ab}_{i}\frac{\delta\Sigma}{\delta{K}^{b}_{\mu}}\right)\nonumber\\
&&=\partial^{2}\partial_{\mu}{\widehat{\mathcal{F}}}^{a}_{\mu i}\,,\\\cr
\Phi^{a}_{i}(\Sigma)&\equiv&\frac{\delta\Sigma}{\delta\phi^{a}_{i}}
+\partial_{\mu}\frac{\delta\S}{\delta{\overline{\mathcal{G}}}^{a}_{\mu i}}
+igf^{abc}\bar\phi^{b}_{i}\frac{\delta\Sigma}{\delta{b}^{c}}
-gf^{abc}\bar{\omega}^{b}_{i}\frac{\delta\Sigma}{\delta\bar{c}^{c}}
+gf^{abc}\frac{\delta\Sigma}{\delta\rho^{bc}_{i}}
-D^{ab}_{\mu}\frac{\delta\S}{\delta{\widehat{\mathcal{H}}}^{b}_{\mu i}}\nonumber\\
&&=-\partial^{2}\partial_{\mu}{\mathcal{G}}^{a}_{\mu i}
+\partial_{\mu}{\mathcal{H}}^{a}_{\mu i}
-\partial_{\mu}{\widehat{\overline{\mathcal{G}}}}^{a}_{\mu i}
-\gamma^{2}gf^{abc}\delta^{c}_{\mu i}A^{b}_{\mu}\,.
\end{eqnarray}

\vspace{3mm}

$\bullet$ the  Ward identities:
\begin{equation}
\mathcal{U}_{i}(\Sigma)=\int d^{4}x\,\biggl\{c^{a}\frac{\delta\Sigma}{\delta\omega^{a}_{i}}
+\bar\omega^{a}_{i}\frac{\delta\Sigma}{\delta\bar{c}^{a}}
-\delta^{ab}\frac{\delta\Sigma}{\delta\rho^{ab}_{i}}\biggr\}=0\,,
\end{equation}
\begin{equation}
\mathcal{V}_{i}(\Sigma)=\int d^{4}x\,\biggl\{-c^{a}\frac{\delta\Sigma}{\delta\phi^{a}_{i}}
+\bar\phi^{a}_{i}\frac{\delta\Sigma}{\delta\bar{c}^{a}}
+\frac{\delta\S}{\delta L^{a}}\frac{\delta\S}{\delta\omega^{a}_{i}}
+\biggl(\frac{\delta\S}{\delta{\widehat{\mathcal{H}}}^{a}_{\mu i}}+\gamma^{2}\delta^{a}_{\mu i}\biggr)\frac{\delta\S}{\delta K^{a}_{\mu}}\biggr\}=0\,.
\end{equation}

\vspace{3mm}

$\bullet$ {the linearly broken $U(4(N^2-1))$ Ward identity}
\begin{eqnarray}
{Q}_{ij}(\Sigma)&\equiv&\int d^{4}x\,\biggl\{\phi^{a}_{i}\frac{\delta\Sigma}{\delta\phi^{a}_{j}}
-\bar\phi^{a}_{j}\frac{\delta\Sigma}{\delta\bar\phi^{a}_{i}}
+\omega^{a}_{i}\frac{\delta\Sigma}{\delta\omega^{a}_{j}}
-\bar\omega^{a}_{j}\frac{\delta\Sigma}{\delta\bar\omega^{a}_{i}}
+{\overline{\mathcal{G}}}^{a}_{\mu i}\frac{\delta\Sigma}{\delta{\overline{\mathcal{G}}}^{a}_{\mu j}}
-{{\mathcal{G}}}^{a}_{\mu j}\frac{\delta\Sigma}{\delta{{\mathcal{G}}}^{a}_{\mu i}}
-{\widehat{\overline{\mathcal{G}}}}^{a}_{\mu j}\frac{\delta\Sigma}{\delta{\widehat{\overline{\mathcal{G}}}}^{a}_{\mu i}}
+{\widehat{\mathcal{G}}}^{a}_{\mu i}\frac{\delta\Sigma}{\delta{\widehat{\mathcal{G}}}^{a}_{\mu j}}\nonumber\\
&&-{\widehat{\overline{\mathcal{F}}}}^{a}_{\mu j}\frac{\delta\Sigma}{\delta{\widehat{\overline{\mathcal{F}}}}^{a}_{\mu i}}
+{\widehat{\mathcal{F}}}^{a}_{\mu i}\frac{\delta\Sigma}{\delta{\widehat{\mathcal{F}}}^{a}_{\mu j}}
-{{{\mathcal{H}}}}^{a}_{\mu j}\frac{\delta\Sigma}{\delta{{{\mathcal{H}}}}^{a}_{\mu i}}
+{\widehat{\mathcal{H}}}^{a}_{\mu i}\frac{\delta\Sigma}{\delta{\widehat{\mathcal{H}}}^{a}_{\mu j}}
+\rho^{ab}_{i}\frac{\delta\Sigma}{\delta\rho^{ab}_{j}}
+\lambda^{ab}_{i}\frac{\delta\Sigma}{\delta\lambda^{ab}_{j}}\biggl\}\nonumber\\
&=&\gamma^{2}\int d^{4}x\,\left(\delta^{a}_{\mu i}\mathcal{H}^{a}_{\mu j}-\delta^{a}_{\mu j}{\widehat{\mathcal{H}}}^{a}_{\mu i}\right)\,.
\label{Qij}
\end{eqnarray}

\vspace{3mm}

$\bullet$ {the exact integrated Ward identities}:
\begin{eqnarray}
T^{(1)}_{ij}(\Sigma)&\equiv&\int d^{4}x\,\left({\overline{\mathcal{F}}}^{a}_{\mu i}\frac{\delta\Sigma}{\delta{\mathcal{G}}^{a}_{\mu j}}
-{\overline{\mathcal{G}}}^{a}_{\mu j}\frac{\delta\Sigma}{\delta{\mathcal{F}}^{a}_{\mu i}}\right)=0\,,\nonumber\cr
T^{(2)}_{ij}(\Sigma)&\equiv&\int d^{4}x\,\left({\widehat{\overline{\mathcal{F}}}}^{a}_{\mu i}\frac{\delta\Sigma}{\delta{\mathcal{G}}^{a}_{\mu j}}
-{\overline{\mathcal{G}}}^{a}_{\mu j}\frac{\delta\Sigma}{\delta{\widehat{\mathcal{F}}}^{a}_{\mu i}}\right)=0\,,\nonumber\cr§ 
T^{(3)}_{ij}(\Sigma)&\equiv&\int d^{4}x\,\left({\widehat{\overline{\mathcal{F}}}}^{a}_{\mu i}\frac{\delta\Sigma}{\delta{\widehat{\mathcal{G}}}^{a}_{\mu j}}
-{\widehat{\overline{\mathcal{G}}}}^{a}_{\mu j}\frac{\delta\Sigma}{\delta{\widehat{\mathcal{F}}}^{a}_{\mu i}}\right)=0\,,\nonumber\cr
T^{(4)}_{ij}(\Sigma)&\equiv&\int d^{4}x\,\left({\overline{\mathcal{F}}}^{a}_{\mu i}\frac{\delta\Sigma}{\delta{\widehat{\mathcal{G}}}^{a}_{\mu j}}
-{\widehat{\overline{\mathcal{G}}}}^{a}_{\mu j}\frac{\delta\Sigma}{\delta{\mathcal{F}}^{a}_{\mu i}}\right)=0\,,\nonumber\cr
T^{(5)}_{ij}(\S)&\equiv&\int d^{4}x\,\left(\mathcal{F}^{a}_{\mu i}\frac{\delta\S}{\delta\mathcal{G}^{a}_{\mu j}}
+{\overline{\mathcal{G}}}^{a}_{\mu j}\frac{\delta\S}{\delta{\overline{\mathcal{F}}}^{a}_{\mu i}}\right)=0\,,\nonumber\cr
T^{(6)}_{ij}(\S)&\equiv&\int d^{4}x\,\left({\overline{\mathcal{F}}}^{a}_{\mu i}\frac{\delta\S}{\delta{\overline{\mathcal{F}}}^{a}_{\mu j}}
-\mathcal{F}^{a}_{\mu j}\frac{\delta\S}{\delta\mathcal{F}^{a}_{\mu i}}\right)=0\,,\nonumber\cr
T^{(7)}_{ij}(\Sigma)&\equiv&(\delta_{ik}\delta_{jl}-\delta_{jk}\delta_{il})\,\int d^{4}x\,{\overline{\mathcal{G}}}^{a}_{\mu k}\frac{\delta\Sigma}{\delta{\mathcal{G}}^{a}_{\mu l}}=0\,,
\nonumber\cr
T^{(8)}_{ij}(\Sigma)&\equiv&\int d^{4}x\,\left({\widehat{\overline{\mathcal{G}}}}^{a}_{\mu i}\frac{\delta\Sigma}{\delta{\mathcal{G}}^{a}_{\mu j}}
-{{\overline{\mathcal{G}}}}^{a}_{\mu j}\frac{\delta\Sigma}{\delta{\widehat{\mathcal{G}}}^{a}_{\mu i}}\right)=0\,,\nonumber\cr
T^{(9)}_{ij}(\Sigma)&\equiv&(\delta_{ik}\delta_{jl}-\delta_{jk}\delta_{il})\,\int d^{4}x\,{\widehat{\overline{\mathcal{G}}}}^{a}_{\mu k}\frac{\delta\Sigma}{\delta{\widehat{\mathcal{G}}}^{a}_{\mu l}}=0\,,
\nonumber\\
T^{(10)}_{ij}(\Sigma)&\equiv&\int d^{4}x\,\left({{\overline{\mathcal{F}}}}^{a}_{\mu i}\frac{\delta\Sigma}{\delta{\overline{\mathcal{F}}}^{a}_{\mu j}}
-{{{\mathcal{F}}}}^{a}_{\mu j}\frac{\delta\Sigma}{\delta{{\mathcal{F}}}^{a}_{\mu i}}\right)=0\,.
\label{Ts}
\end{eqnarray}

\vspace{3mm}

$\bullet$ {the $SL(2,\mathds{R})$ Ward identity}
\begin{equation}
\mathcal{D}(\Sigma)\equiv\int d^{4}x\,\left(c^{a}\frac{\delta\Sigma}{\delta\bar{c}^{a}}
-i\frac{\delta\Sigma}{\delta{b}^{a}}\frac{\delta\Sigma}{\delta{L}^{a}}\right)=0\,.
\end{equation}

\vspace{3mm}

$\bullet$ {the linearly broken rigid $SU(N)$ symmetry}
\begin{equation}
\mathcal{W}^{a}(\Sigma)=-\gamma^{2}\int d^{4}x\,gf^{abc}(\mathcal{H}^{b}_{\mu i}\delta^{c}_{\mu i}+{\widehat{\mathcal{H}}}^{b}_{\mu i}\delta^{c}_{\mu i})\,,
\end{equation}
with
\begin{eqnarray}
\mathcal{W}^{a}&\equiv&gf^{abc}\int d^{4}x\,\Biggl\{\,\sum_{y\in\mathcal{O}}y^{b}\frac{\delta}{\delta y^{c}}+\rho^{bd}_{i}\frac{\delta}{\delta\rho^{cd}_{i}}
+\rho^{db}_{i}\frac{\delta}{\delta\rho^{dc}_{i}}
+\lambda^{bd}_{i}\frac{\delta}{\delta\lambda^{cd}_{i}}
+\lambda^{db}_{i}\frac{\delta}{\delta\lambda^{dc}_{i}}\Biggr\}
\end{eqnarray}
where $\mathcal{O}$ stands for
\begin{equation}
\mathcal{O}=\left\{A^{a}_{\mu}, b^{a}, \bar{c}^{a},c^{a},\phi^{a}_{i},\bar\phi^{a}_{i},\omega^{a}_{i},\bar\omega^{a}_{i}, \dots\right\}\;,
\end{equation}
{\it i.e.}, the set $\mathcal{O}$ is the set of all fields and sources  that have only one color index, where we have not taken into account the color index  hidden in the multi-index $i=(a,\mu)$.

\vspace{3mm}

$\bullet$ {the equation of motion of the source $\lambda^{ab}_{i}$}
\begin{equation}
\Lambda^{ab}_{i}(\Sigma)\equiv\frac{\delta\Sigma}{\delta\lambda^{ab}_{i}}-\left(\frac{\delta\Sigma}{\delta{\widehat{\mathcal{H}}}^{a}_{\mu i}}+\gamma^{2}\delta^{a}_{\mu i}\right)\frac{\delta\Sigma}{\delta{K}^{b}_{\mu}}=0\,.
\end{equation}
\vspace{3mm}

$\bullet$ {the $Q_f$ charge}

\noindent We can combine the operators $Q_{ij}$ and $T^{(6)}_{ij}$, appearing in eqs.\eqref{Qij} and \eqref{Ts}, respectively, and construct the following operator:
\begin{equation}
Q^{T}_{ij}=Q_{ij}+ T^{(6)}_{ij}\,.
\end{equation}
The operator $Q^{T}_{ij}$ commutes with the BRST operator $s$
\begin{equation}
[s,Q^{T}_{ij}]=0 \,.
\end{equation}
Then,  the trace of $Q^{T}_{ij}$ defines a new charge:
\begin{eqnarray}
Q^{T}_{ii}\equiv Q_{f}&\!\!\!\!\!\!:=\!\!\!\!\!\!\!\!&\int d^{4}x\,\biggl\{\phi^{a}_{i}\frac{\delta}{\delta\phi^{a}_{i}}
-\bar\phi^{a}_{i}\frac{\delta}{\delta\bar\phi^{a}_{i}}
+\omega^{a}_{i}\frac{\delta}{\delta\omega^{a}_{i}}
-\bar\omega^{a}_{i}\frac{\delta}{\delta\bar\omega^{a}_{i}}
+{\overline{\mathcal{G}}}^{a}_{\mu i}\frac{\delta}{\delta{\overline{\mathcal{G}}}^{a}_{\mu i}}
-{{\mathcal{G}}}^{a}_{\mu i}\frac{\delta}{\delta{{\mathcal{G}}}^{a}_{\mu i}}
-{\widehat{\overline{\mathcal{G}}}}^{a}_{\mu i}\frac{\delta}{\delta{\widehat{\overline{\mathcal{G}}}}^{a}_{\mu i}}
+{\widehat{\mathcal{G}}}^{a}_{\mu i}\frac{\delta}{\delta{\widehat{\mathcal{G}}}^{a}_{\mu i}}\nonumber\\
&&-{\widehat{\overline{\mathcal{F}}}}^{a}_{\mu i}\frac{\delta}{\delta{\widehat{\overline{\mathcal{F}}}}^{a}_{\mu i}}
+{\widehat{\mathcal{F}}}^{a}_{\mu i}\frac{\delta}{\delta{\widehat{\mathcal{F}}}^{a}_{\mu i}}
-{{{\mathcal{H}}}}^{a}_{\mu i}\frac{\delta}{\delta{{{\mathcal{H}}}}^{a}_{\mu i}}
+{\widehat{\mathcal{H}}}^{a}_{\mu i}\frac{\delta}{\delta{\widehat{\mathcal{H}}}^{a}_{\mu i}}
+\rho^{ab}_{i}\frac{\delta}{\delta\rho^{ab}_{i}}
+\lambda^{ab}_{i}\frac{\delta}{\delta\lambda^{ab}_{i}}
+{\overline{\mathcal{F}}}^{a}_{\mu i}\frac{\delta}{\delta{\overline{\mathcal{F}}}^{a}_{\mu i}}
-\mathcal{F}^{a}_{\mu i}\frac{\delta}{\delta\mathcal{F}^{a}_{\mu i}}\biggl\}\,,\nonumber\\
\end{eqnarray}
where $f\equiv4(N^{2}-1)$. The $Q_f$ charge gives rise to a powerful  linearly broken Ward identity when acting on $\S$, namely
\begin{equation}
Q_f(\S)=\gamma^{2}\int d^{4}x\,\left(\delta^{a}_{\mu i}\mathcal{H}^{a}_{\mu i}-\delta^{a}_{\mu i}{\widehat{\mathcal{H}}}^{a}_{\mu i}\right)\,.  \label{qfid}
\end{equation}
This is actually the Ward identity which enables us to make use of the multi-index $i=(a,\mu)$.

\section{Proof of the all orders renormalizability}
Having established the Ward identities obeyed by the action $\Sigma$, eqs.\eqref{sti}-\eqref{qfid}, we can proceed to show the renormalizability to all orders of the model. Let us begin with the algebraic characterization of the most general local invariant counterterm that is compatible with all Ward identities.

\renewcommand{\d}{\delta}

\subsection{Algebraic characterization of the invariant  counterterm}

In order to characterize  the most general local invariant counterterm which can be freely added to all orders in perturbation theory, we follow the general setup of the algebraic renormalization \cite{Piguet:1995er} and perturb the starting action $\S$ by adding an integrated local polynomial in the fields and sources, $\S_{count}$,
with dimension bounded by four and with vanishing ghost number. We thus demand that the perturbed action,
\begin{equation}
\S+\eta\,\S_{count}\,,
\end{equation}
where $\eta$ is an expansion parameter, fulfills, to the first order in $\eta$, the same set of Ward identities obeyed by $\S$, eqs.\eqref{sti}-\eqref{qfid}. This requirement gives rise to the following constraints for the counterterm $\S_{count}$:
\begin{equation}
\mathcal{S}_{\S}(\S_{count})=0\,,
\label{linearized_count}
\end{equation}
\begin{equation}
\left(\frac{\delta}{\delta\bar{c}^{a}}+\partial_{\mu}\frac{\delta}{\delta K^{a}_{\mu}}\right)\S_{count}=0\,,
\label{antigh_count}
\end{equation}
\begin{eqnarray}
&\displaystyle \frac{\d}{\d b^{a}}\S_{count}=0\,,\qquad\frac{\partial}{\partial \gamma^{2}}\S_{count}=0\,,\qquad
\frac{\d}{\d{\mathcal{G}}^{a}_{\mu i}}\S_{count}=0\,,\qquad
\frac{\d}{\d{\widehat{\mathcal{G}}}^{a}_{\mu i}}\S_{count}=0\,,\qquad
\frac{\d}{\d{\mathcal{F}}^{a}_{\mu i}}\S_{count}=0\,,&\nonumber\\
&\displaystyle\frac{\d}{\d{\overline{\mathcal{F}}}^{a}_{\mu i}}\S_{count}=0\,,\qquad
\frac{\d}{\d{\widehat{\mathcal{F}}}^{a}_{\mu i}}\S_{count}=0\,,\qquad
\frac{\d}{\d{\mathcal{H}}^{a}_{\mu i}}\S_{count}=0\,,\qquad
\frac{\d}{\d{\widehat{\mathcal{H}}}^{a}_{\mu i}}\S_{count}=0\,,&
\label{independent}
\end{eqnarray}
\begin{eqnarray}
&\overline{\Phi}^{a}_{i}(\S_{count})=0\,,\qquad\overline{\Omega}^{a,i}_{\S}(\S_{count})=0\,,\qquad\Phi^{a}_{i}(\S_{count})=0\,,
\qquad\mathcal{U}_{i}(\S_{count})=0\,,&\nonumber\\
&\mathcal{V}^{i}_{\S}(\S_{count})=0\,,\qquad
\mathcal{D}_{\S}(\S_{count})=0\,,\qquad
\mathcal{W}^{a}(\S_{count})=0\,,\qquad
\Lambda^{ab,i}_{\S}(\S_{count})=0\,,&\nonumber\\
&Q_{ij}(\S_{count})=0\,,\qquad T_{ij}^{(n)}(\S_{count})=0\,,\qquad n=1,\dots,10\,,&
\end{eqnarray}
\begin{equation}
Q_{f}(\S_{count})=0\,.
\label{Qf_count}
\end{equation}
Here, the operators with the subscript ``$\S$'' represent the so called linearized operators corresponding to the Ward identities which are nonlinear in $\Sigma$, see \cite{Piguet:1995er}. For example,  $\mathcal{S}_{\S}$ is the linearized operator corresponding to the Slavnov-Taylor identity  \eqref{sti}, namely
\begin{eqnarray}
\mathcal{S}_\Sigma&=&\int d^{4}x\,\biggl\{
\frac{\delta\Sigma}{\delta{K}^{a}_{\mu}}\frac{\delta}{\delta{A}^{a}_{\mu}}
+\frac{\delta\Sigma}{\delta{A}^{a}_{\mu}}\frac{\delta}{\delta{K}^{a}_{\mu}}
+\frac{\delta\Sigma}{\delta{L}^{a}}\frac{\delta}{\delta{c}^{a}}
+\frac{\delta\Sigma}{\delta{c}^{a}}\frac{\delta}{\delta{L}^{a}}
+ib^{a}\frac{\delta}{\delta\bar{c}^{a}}
+\bar\phi^{a}_{i}\frac{\delta}{\delta\bar{\omega}^{a}_{i}}
+\omega^{a}_{i}\frac{\delta}{\delta{\phi}^{a}_{i}}\nonumber\\
&&+\overline{\mathcal{G}}^{a}_{\mu i}\frac{\delta}{\delta{\overline{\mathcal{F}}}^{a}_{\mu i}}
+{\mathcal{F}}^{a}_{\mu i}\frac{\delta}{\delta{{\mathcal{G}}}^{a}_{\mu i}}
+{\widehat{\overline{\mathcal{G}}}}^{a}_{\mu i}\frac{\delta}{\delta{\widehat{\overline{\mathcal{F}}}}^{a}_{\mu i}}
+{\widehat{\mathcal{F}}}^{a}_{\mu i}\frac{\delta}{\delta{\widehat{{\mathcal{G}}}}^{a}_{\mu i}}
+\rho^{ab}_{i}\frac{\delta}{\delta\lambda^{ab}_{i}}\biggr\}\,.
\end{eqnarray}
As the BRST operator, also $\mathcal{S}_{\S}$ is nilpotent, {\it i.e.}
\begin{equation}
\mathcal{S}_{\S} \mathcal{S}_{\S} = 0 \;.
\end{equation}
The remaining linearized operators are given by:
\begin{eqnarray}
\overline{\Omega}^{a,i}_{\S}&=&
\frac{\delta}{\delta\bar\omega^{a}_{i}}
+\partial_{\mu}\frac{\delta}{\delta{\widehat{\overline{\mathcal{F}}}}^{a}_{\mu i}}
-gf^{abc}\left(\frac{\delta\Sigma}{\delta\mathcal{H}^{b}_{\mu i}}+\delta^{b}_{\mu i}\gamma^{2}\right)\frac{\delta}{\delta{K}^{c}_{\mu}}
+gf^{abc}\frac{\delta\Sigma}{\delta{K}^{b}_{\mu}}\frac{\delta}{\delta\mathcal{H}^{c}_{\mu i}}
+\partial_{\mu}\left(\rho^{ab}_{i}\frac{\delta}{\delta{K}^{b}_{\mu}}\right)\,,\nonumber\\
\mathcal{V}^{i}_{\Sigma}&=&\int d^{4}x\,\biggl\{-c^{a}\frac{\delta}{\delta\phi^{a}_{i}}
+\bar\phi^{a}_{i}\frac{\delta}{\delta\bar{c}^{a}}
+\frac{\d\S}{\d L^{a}}\frac{\d}{\d\omega^{a}_{i}}
+\frac{\d\S}{\d\omega^{a}_{i}}\frac{\d}{\d L^{a}}
+\biggl(\frac{\d\S}{\d{\widehat{\mathcal{H}}}^{a}_{\mu i}}+\gamma^{2}\d^{a}_{\mu i}\biggr)\frac{\d}{\d K^{a}_{\mu}}
+\frac{\d\S}{\d K^{a}_{\mu}}\frac{\d}{\d{\widehat{\mathcal{H}}}^{a}_{\mu i}}\biggr\}\,,\nonumber\\
\mathcal{D}_{\Sigma}&=&\int d^{4}x\,\left(c^{a}\frac{\delta}{\delta\bar{c}^{a}}
-i\frac{\delta\Sigma}{\delta{b}^{a}}\frac{\delta}{\delta{L}^{a}}
-i\frac{\delta\Sigma}{\delta{L}^{a}}\frac{\delta}{\delta{b}^{a}}\right)\,,\nonumber\\
\Lambda^{ab,i}_{\S}&=&\frac{\delta}{\delta\lambda^{ab}_{i}}-\left(\frac{\delta\Sigma}{\delta{\widehat{\mathcal{H}}}^{a}_{\mu i}}
+\gamma^{2}\delta^{a}_{\mu i}\right)\frac{\delta}{\delta{K}^{b}_{\mu}}
-\frac{\delta\Sigma}{\delta{K}^{b}_{\mu}}\frac{\delta}{\delta{\widehat{\mathcal{H}}}^{a}_{\mu i}}\,.
\end{eqnarray}
Let us now turn to the characterization of  the counterterm. The constraints \eqref{independent} imply that  $\Sigma_{count}$ is independent from  the fields $b$, $\mathcal{G}$, $\widehat{\mathcal{G}}$,
$\mathcal{F}$, $\widehat{\mathcal{F}}$, $\overline{\mathcal{F}}$, $\mathcal{H}$, $\widehat{\mathcal{H}}$, as well as from  the Gribov parameter $\gamma^{2}$. Equation \eqref{antigh_count} means that  $\S_{count}$ depends on $\bar{c}$ and $K$ only through the combination $(\partial_{\mu}\bar{c}^{a}+K^{a}_{\mu})$. Moreover, from eq.\eqref{Qf_count} it follows that  $\S_{count}$ has zero $Q_f$-charge. Finally, relying on well known properties of the cohomology of Yang-Mills theories \cite{Piguet:1995er}, condition \eqref{linearized_count} allow us to construct the
countertem in the form:
\begin{eqnarray}
\Sigma_{count}& = & \; a_0\,S_{\mathrm{YM}}+\mathcal{S}_{\Sigma}\Delta^{(-1)} \,,  \nonumber \\[3mm]
S_{\mathrm{YM}} & = & \int d^4x \; \frac{1}{4} F^a_{\mu\nu} F^a_{\mu\nu}   \;, \label{ct}
\end{eqnarray}
where $a_0$ is a dimensionless coefficient and  $\Delta^{(-1)}$ is an integrated polynomial in the fields and sources with dimension four and ghost number $-1$.  Collecting all this information, and making use of Table 1 and of Table 2, one can write $\Delta^{(-1)}$ as
\begin{eqnarray}\label{mon}
\Delta^{(-1)}&=&\int d^{4}x\,\biggl\{
a_{1}\,(\partial_{\mu}\bar{c}^{a}+K^{a}_{\mu})A^{a}_{\mu}
+a_{2}\,L^{a}c^{a}
+a_{3}\,{\widehat{\overline{\mathcal{F}}}}^{a}_{\mu i}\,\partial_{\mu}\phi^{a}_{i}
+a_{4}\,gf^{abc}{\widehat{\overline{\mathcal{F}}}}^{a}_{\mu i}\,A^{c}_{\mu}\phi^{b}_{i}
+a_{5}\,\bar\omega^{a}_{i}\,\partial^{2}\phi^{a}_{i}
+a_{6}\,gf^{abc}(\partial_{\mu}\bar\omega^{a}_{i})A^{c}_{\mu}\phi^{b}_{i}\nonumber\\
&&+a_{7}\,gf^{abc}\bar\omega^{a}_{i}A^{c}_{\mu}\,\partial_{\mu}\phi^{b}_{i}
+a_{8}\,gf^{abc}\bar\omega^{a}_{i}A^{c}_{\mu}{\overline{\mathcal{G}}}^{b}_{\mu i}
+a_{9}\,\bar\omega^{a}_{i}\,\partial_{\mu}{\overline{\mathcal{G}}}^{a}_{\mu i}
+a_{10}\,{\widehat{\overline{\mathcal{F}}}}^{a}_{\mu i}{\overline{\mathcal{G}}}^{a}_{\mu i}
+t_{1}^{abcd}\,\bar\omega^{a}_{i}\phi^{b}_{i}\bar\phi^{c}_{j}\phi^{d}_{j}
+t_{2}^{abcd}\,\bar\omega^{a}_{i}\phi^{b}_{j}\bar\phi^{c}_{i}\phi^{d}_{j}\nonumber\\
&&
+t_{3}^{abcd}\,\bar\omega^{a}_{i}\phi^{b}_{i}\bar\omega^{c}_{j}\omega^{d}_{j}
+t_{4}^{abcd}\,\bar\omega^{a}_{i}\phi^{b}_{j}\bar\omega^{c}_{i}\omega^{d}_{j}
+\alpha_{1}^{abcd}\,\lambda^{ab}_{i}{\widehat{\overline{\mathcal{F}}}}^{c}_{\mu i}\,\partial_{\mu}c^{d}
+\alpha_{2}^{abcd}\,\lambda^{ab}_{i}\,(\partial_{\mu}{\widehat{\overline{\mathcal{F}}}}^{c}_{\mu i})\,c^{d}
+\alpha_{3}^{abcd}\,\lambda^{ab}_{i}{\widehat{\overline{\mathcal{G}}}}^{c}_{\mu i}\,A^{d}_{\mu}\nonumber\\
&&+\alpha_{4}^{abcd}\,\rho^{ab}_{i}{\widehat{\overline{\mathcal{F}}}}^{c}_{\mu i}\,A^{d}_{\mu}
+\alpha_{5}^{abcd}\lambda^{ab}_{i}(\partial_{\mu}\bar\omega^{c}_{i})\,\partial_{\mu}c^{d}
+\alpha_{6}^{abcd}\,\lambda^{ab}_{i}(\partial^{2}\bar\omega^{c}_{i})c^{d}
+\alpha_{7}^{abcd}\,\lambda^{ab}_{i}\bar\omega^{c}_{i}\,\partial^{2}c^{d}
+\alpha_{8}^{abcd}\,(\partial^{2}\lambda^{ab}_{i})\bar\omega^{c}_{i}c^{d}\nonumber\\
&&+\alpha_{9}^{abcd}\,\lambda^{ab}_{i}(\partial_{\mu}\bar\phi^{c}_{i})A^{d}_{\mu}
+\alpha_{10}^{abcd}\,\lambda^{ab}_{i}\bar\phi^{c}_{i}\,\partial_{\mu}A^{d}_{\mu}
+\alpha_{11}^{abcd}\,\rho^{ab}_{i}(\partial_{\mu}\bar\omega^{c}_{i})A^{d}_{\mu}
+\alpha_{12}^{abcd}\,\rho^{ab}_{i}\bar\omega^{c}_{i}\,\partial_{\mu}A^{d}_{\mu}
+\beta_{1}^{abcde}\,\lambda^{ab}_{i}{\widehat{\overline{\mathcal{F}}}}^{c}_{\mu i}c^{d}A^{e}_{\mu}\nonumber\\
&&+\beta_{2}^{abcde}\,\lambda^{ab}_{i}(\partial_{\mu}\bar\omega^{c}_{i})c^{d}A^{e}_{\mu}
+\beta_{3}^{abcde}\,\lambda^{ab}_{i}\bar\omega^{c}_{i}(\partial_{\mu}c^{d})A^{e}_{\mu}
+\beta_{4}^{abcde}\,\lambda^{ab}_{i}\bar\omega^{c}_{i}c^{d}(\partial_{\mu}A^{e}_{\mu})
+\beta_{5}^{abcde}\,\lambda^{ab}_{i}\bar\phi^{c}_{i}A^{d}_{\mu}A^{e}_{\mu}\nonumber\\
&&+\beta_{6}^{abcde}\,\rho^{ab}_{i}\bar\omega^{c}_{i}A^{d}_{\mu}A^{e}_{\mu}
+\tau^{abcdef}\,\lambda^{ab}_{i}\bar\omega^{c}_{i}c^{d}A^{e}_{\mu}A^{f}_{\mu}
+M_{1}^{abcdefg}\,\lambda^{ab}_{i}\lambda^{cd}_{j}\bar\phi^{e}_{i}\bar\phi^{f}_{j}c^{g}
+M_{2}^{abcdefg}\,\lambda^{ab}_{i}\lambda^{cd}_{i}\bar\phi^{e}_{j}\bar\phi^{f}_{j}c^{g}\biggr\}\,.
\end{eqnarray}
In this expression, $a_i, i=1,...,10$, are dimensionless coefficients, while $\{t\}$, $\{\alpha\}$,  $\{\beta\}$, $\{\tau\}$, $\{M\}$  stand for invariant tensors of the gauge group $SU(N)$.  Following an observation already employed in previous works \cite{Dudal:2007cw,Dudal:2008sp,Dudal:2011gd}, it turns out that the coefficient $a_2$ vanishes. This is due to the fact that, as the term $L^a c^a$ is already of dimension 4, the coefficient $a_2$ cannot depend on the Gribov parameter $\gamma^2$, and it vanishes when $\gamma^2=0$ due to the existence of an additional Ward identity, called the Landau gauge ghost Ward identity, see  \cite{Blasi:1990xz,Piguet:1995er}.

\noindent Furthermore, applying the remaining constraints, and using the following useful commutation
and anti-commutation relations,
\begin{eqnarray}
&\left[\mathcal{S}_{\Sigma},\frac{\delta}{\delta{b}^{a}}\right]=-i\left(\frac{\delta}{\delta\bar{c}^{a}}+\partial_{\mu}\frac{\delta}{\delta{K}^{a}_{\mu}}\right)\,,\qquad
\left\{\mathcal{S}_{\Sigma},\frac{\delta}{\delta{\mathcal{F}}^{a}_{\mu i}}\right\}=\frac{\delta}{\delta{\mathcal{G}}^{a}_{\mu i}}\,,\qquad
\left\{\mathcal{S}_{\Sigma},\frac{\delta}{\delta{\widehat{\mathcal{F}}}^{a}_{\mu i}}\right\}=\frac{\delta}{\delta{\widehat{\mathcal{G}}}^{a}_{\mu i}}\,,&\nonumber\\\cr
&\left\{\mathcal{S}_{\Sigma},\frac{\delta}{\delta{\overline{\mathcal{F}}}^{a}_{\mu i}}\right\}=0\,,\qquad
\left[\mathcal{S}_{\Sigma},\frac{\delta}{\delta{\mathcal{G}}^{a}_{\mu i}}\right]=0\,,\qquad
\left[\mathcal{S}_{\Sigma},\frac{\delta}{\delta{\widehat{\mathcal{G}}}^{a}_{\mu i}}\right]=0\,,&\nonumber\\\cr
&\left[\mathcal{S}_{\Sigma},\overline{\Phi}^{a}_{i}\right]=-\overline{\Omega}^{ai}_{\Sigma}\,,\qquad
\left\{\mathcal{S}_{\Sigma},\overline{\Omega}^{ai}_{\Sigma}\right\}=0\,,\qquad
\left[\mathcal{S}_{\Sigma},\Phi^{a}_{i}\right]=-gf^{abc}\Lambda^{bc,i}_{\Sigma}\,,&\nonumber\\\cr
&\left[\mathcal{S}_{\Sigma},\mathcal{U}_{i}\right]=\displaystyle{\int }d^{4}x\,\left(\mathcal{V}^{i}_{\S}+\delta^{ab}\Lambda^{ab,i}_{\Sigma}\right)\,,\qquad
\left\{\mathcal{S}_{\Sigma},\mathcal{V}^{i}_{\S}\right\}=0\,,\qquad
\left\{\mathcal{S}_{\Sigma},\Lambda^{ab,i}_{\Sigma}\right\}=0\,,&\nonumber\\\cr
&\left\{\mathcal{S}_{\Sigma},T^{(1)}_{ij}\right\}=T^{(7)}_{ij}\,,\qquad
\left\{\mathcal{S}_{\Sigma},T^{(2)}_{ij}\right\}=T^{(8)}_{ij}\,,\qquad
\left\{\mathcal{S}_{\Sigma},T^{(3)}_{ij}\right\}=T^{(9)}_{ij}\,,&\nonumber\\\cr
&\left\{\mathcal{S}_{\Sigma},T^{(4)}_{ij}\right\}=-T^{(7)}_{ji}\,,\qquad
\left[\mathcal{S}_{\Sigma},T^{(10)}_{ij}\right]=0\,,&\nonumber\\\cr
&\left[\mathcal{S}_{\Sigma},\mathcal{D}_{\Sigma}\right]=0\,,\qquad
\left[\mathcal{S}_{\Sigma},\mathcal{W}^{a}\right]=0\,,\qquad
\left[\mathcal{S}_{\Sigma},Q^T_{ij}\right]=0 \,,&
\end{eqnarray}
it follows that, after a lengthy analysis, only the coefficient $a_1$ remains free. The a priori quite monstrous expression for $\Delta^{(-1)}$, eq.~\eqref{mon}, eventually thus reduces considerably to the following form
\begin{eqnarray}
\Delta^{(-1)}&=&a_{1}\int d^{4}x\left[(\partial_{\mu}\bar{c}^{a}+K^{a}_{\mu})A^{a}_{\mu}
+(\partial_{\mu}\bar\omega^{a}_{i}+{\widehat{\overline{\mathcal{F}}}}^{a}_{\mu i})D^{ab}_{\mu}\phi^{b}_{i}
+{\overline{\mathcal{G}}}^{a}_{\mu i}\,D^{ab}_{\mu}\bar\omega^{b}_{i}
+\,{\widehat{\overline{\mathcal{F}}}}^{a}_{\mu i}{\overline{\mathcal{G}}}^{a}_{\mu i}
-\lambda^{ab}_{i}\,{\widehat{\overline{\mathcal{F}}}}^{a}_{\mu i}\,D^{bc}_{\mu}c^{c}\right]\,.
\end{eqnarray}
Summarizing, the most general invariant counterterm $\Sigma_{count}$ compatible with all constraints \eqref{linearized_count}--\eqref{Qf_count} has two independent free coefficients, $a_0, a_1$, and is given by
\begin{equation}
\Sigma_{count} = a_0  \int d^4x \; \frac{1}{4} F^a_{\mu\nu} F^a_{\mu\nu}\; +\;\mathcal{S}_{\Sigma}\Delta^{(-1)} \;. \label{fct}
\end{equation}

\begin{table}
\begin{center}
\begin{tabular}{c|c|c|c|c|c|c|c|c|c|c|c|c|c|c|c|c|c|c}
\hline\hline
&$A$&$b$&$\bar{c}$&$c$&$\bar\phi$&$\phi$&$\bar\omega$&$\omega$&$\overline{\mathcal{G}}$&$\mathcal{G}$&$\overline{\mathcal{F}}$&$\mathcal{F}$
&$\widehat{\overline{\mathcal{G}}}$&$\widehat{\mathcal{G}}$&$\widehat{\overline{\mathcal{F}}}$&$\widehat{\mathcal{F}}$&$\mathcal{H}$&$\widehat{\mathcal{H}}$\cr
\hline
dimension&1&2&2&0&1&1&1&1&2&0&2&0&2&0&2&0&2&2\cr
ghost $\#$&$0$&$0$&$-1$&$1$&$0$&$0$&$-1$&$1$&$0$&$0$&$-1$&$1$&$0$&$0$&$-1$&$1$&$0$&$0$\cr
$Q_{f}$-charge&$0$&$0$&$0$&$0$&$-1$&$1$&$-1$&$1$&$1$&$-1$&$1$&$-1$&$-1$&$1$&$-1$&$1$&$-1$&$1$\cr
\hline\hline
\end{tabular}
\caption{Quantum numbers of the fields}
\end{center}
\label{table1}
\end{table}
\begin{table}
\begin{center}
\begin{tabular}{c|c|c|c|c}
\hline\hline
&$K$&$L$&$\rho$&$\lambda$\cr
\hline
dimension&3&4&1&1\cr
ghost $\#$&$-1$&$-2$&$0$&$-1$\cr
$Q_{f}$-charge&$0$&$0$&$1$&$1$\cr
\hline\hline
\end{tabular}
\caption{Quantum numbers of the external sources}
\end{center}
\label{table2}
\end{table}

\subsection{Renormalization factors}
Having characterized the most general invariant local counterterm $\Sigma_{count}$, eq.\eqref{fct}, compatible with all Ward identities \eqref{sti}-\eqref{qfid}, it remains to check if $\Sigma_{count}$ can be reabsorbed into the starting action $\Sigma$ through a multiplicative renormalization of the fields, sources and parameters of theory, namely
\begin{equation}
\Sigma(f,J) \;+\; \eta \Sigma_{count}(f,J) = \Sigma(f_0,J_0) + O(\eta^2) \;, \label{renorm}
\end{equation}
with
\begin{equation}
f_{0}=Z^{1/2}_{f}\,f\,,\qquad J_{0}=Z_{J}\,J\,,
\end{equation}
where $(f_0, f)$ is a shorthand notation for  the bare and renormalized fields, while $(J_0,J)$ stand for  the bare and renormalized sources and parameters. Making use of expression \eqref{fct}, for the renormalization factors $\{ Z\}$ one
obtains
\begin{eqnarray}
Z_{A}^{1/2}&=&1+\eta\left(\frac{a_{0}}{2}+a_{1}\right)+O(\eta^{2})\,,\nonumber\\
Z_{g}&=&1-\eta\,\frac{a_{0}}{2}+O(\eta^{2})\,,    \label{zz}
\end{eqnarray}
\begin{eqnarray}
&&Z^{1/2}_{c}=Z^{1/2}_{\bar{c}}=Z_{\phi}^{1/2}=Z_{\bar{\phi}}^{1/2}=Z_{g}^{-1/2}Z_{A}^{-1/4}\,,\nonumber\\
&&Z_{\bar\omega}^{1/2}=Z^{-1}_{g}\,,\qquad Z_{\omega}^{1/2}=Z_{A}^{-1/2}\,,\nonumber\\
&&Z^{1/2}_{\overline{\mathcal{G}}}=Z^{1/2}_{{\widehat{\overline{\mathcal{G}}}}}=Z_{\gamma^{2}}=Z_{g}^{-1/2}Z_{A}^{-1/4}\,,\nonumber\\
&&Z^{1/2}_{{\mathcal{G}}}=Z^{1/2}_{{\widehat{{\mathcal{G}}}}}=Z_{g}^{1/2}Z_{A}^{1/4}\,,\nonumber\\
&&Z^{1/2}_{{\widehat{\overline{\mathcal{F}}}}}=Z_{g}^{-1}\,,\qquad Z^{1/2}_{{\widehat{{\mathcal{F}}}}}=Z_{g}\,,\nonumber\\
&&Z^{1/2}_{{{\overline{\mathcal{F}}}}}=Z^{1/2}_{{{{\mathcal{F}}}}}=1\,,\nonumber\\
&&Z^{1/2}_{{{{\mathcal{H}}}}}= Z^{1/2}_{{\widehat{{\mathcal{H}}}}}=Z_{g}^{1/2}Z^{1/4}_{A}\,,\nonumber\\
&&Z_{K}=Z_{g}^{1/2}Z^{1/4}_{A}\,,\qquad Z_{L}=Z_{g}Z^{1/2}_{A}\,,\nonumber\\
&&Z_{\rho}=Z_{g}^{3/2}Z^{1/4}_{A}\,,\qquad Z_{\lambda}=Z_{g}^{-1}Z^{-1/2}_{A}\,.
\end{eqnarray}
This finalizes the proof of the all orders algebraic renormalization of the novel formulation of the Gribov-Zwanziger theory. Let us conclude by observing that the renormalization factor $Z_{\gamma^{2}}$ of the Gribov parameter $\gamma^2$ is not an independent quantity, being expressed in terms of the renormalization factors of the gauge coupling constant $g$ and of the gauge field $A^a_\mu$,  {\it i.e.}  $  Z_{\gamma^{2}}=Z_{g}^{-1/2}Z_{A}^{-1/4}$. This feature expresses the nonrenormalization properties of $\gamma^2$, already established in  \cite{Zwanziger:1989mf,Zwanziger:1992qr,Dudal:2007cw,Dudal:2008sp,Dudal:2011gd}, here a simple consequence of the powerful Ward identity \eqref{ge}.
\section{Conclusion}
In this work we have pursued the investigation of the novel formulation of the Gribov-Zwanziger action proposed in  \cite{Dudal:2012sb}, which allows for an exact BRST invariance of the action implementing the restriction to the Gribov horizon. As shown in  \cite{Dudal:2012sb}, the BRST symmetry turns out to be spontaneously broken, the breaking parameter being nothing but the Gribov mass $\gamma^2$. It is worth mentioning that in this reformulation the BRST operator $s$ does keep its nilpotency, {\it i.e.} $s^2=0$, a crucial feature which enables us to employ the powerful results on the cohomology of $s$ in order to construct the set of colorless local composite gauge invariant operators \cite{Piguet:1995er}.

\noindent In the present paper we have presented the all orders algebraic proof of the renormalizability of the new formulation. In particular, as one can see from eq.~\eqref{zz}, only two independent renormalization factors are needed, namely $Z_{A}$ and $Z_{g}$, a feature which is shared by the original formulation of the Gribov-Zwanziger action. This is an important check of the equivalence between the two formulations at the quantum level.

\noindent Certainly, many aspects of the role of the BRST symmetry in the presence of the Gribov horizon remain to be unraveled. Though, we believe that the current formulation in which the BRST symmetry is spontaneously broken might be helpful in order to face the hard and still open problem of identifying a set of renormalizable composite operators whose correlation functions display the necessary analytical and unitarity properties allowing to make contact with  the physical spectrum of a confining Yang-Mills theory. The results of \cite{Dudal:2012sb}, together with those of the current follow-up paper already learn that we can introduce the subspace of renormalizable gauge invariant operators which is furthermore preserved under time evolution, based on BRST cohomology tools. The further extraction of a physical subspace with the desired spectral properties is now subject to further investigation.

\noindent Finally, although the proof of the renormalizability given here refers to the new Gribov-Zwanziger action in $4d$, it is worth to mention that it immediately generalizes to the case of the Refined Gribov-Zwanziger action \cite{Dudal:2007cw,Dudal:2008sp,Dudal:2011gd}, both in $4d$ and in $3d$ \cite{Dudal:2008rm}.

\section*{Acknowledgments}
The Conselho Nacional de Desenvolvimento Cient\'{\i}fico e
Tecnol\'{o}gico (CNPq-Brazil), the Faperj, Funda{\c{c}}{\~{a}}o de
Amparo {\`{a}} Pesquisa do Estado do Rio de Janeiro, the SR2-UERJ,  the
Coordena{\c{c}}{\~{a}}o de Aperfei{\c{c}}oamento de Pessoal de
N{\'{\i}}vel Superior (CAPES)  are gratefully acknowledged. D.~D.~is supported by the Research-Foundation Flanders. L.~F.~P. is supported by an Alexander von Humboldt Foundation fellowship.



\begin{thebibliography}{99}


\bibitem{Dudal:2012sb}
  D.~Dudal and S.~P.~Sorella,
  Phys.\ Rev.\ D {\bf 86}, 045005 (2012).

\bibitem{Gribov:1977wm}
  V.~N.~Gribov,
  Nucl.\ Phys.\ B {\bf 139}, 1 (1978).


\bibitem{Sobreiro:2005ec}
  R.~F.~Sobreiro and S.~P.~Sorella,
  hep-th/0504095.


\bibitem{Vandersickel:2012tz}
  N.~Vandersickel and D.~Zwanziger,
  Phys.\ Rept.\  {\bf 520}, 175 (2012).

\bibitem{Singer:1978dk}
  I.~M.~Singer,
  Commun.\ Math.\ Phys.\  {\bf 60}, 7 (1978).

\bibitem{Zwanziger:1989mf}
  D.~Zwanziger,
  Nucl.\ Phys.\ B {\bf 323}, 513 (1989).

\bibitem{Zwanziger:1992qr}
  D.~Zwanziger,
  Nucl.\ Phys.\ B {\bf 399}, 477 (1993).


\bibitem{Dudal:2007cw}
  D.~Dudal, S.~P.~Sorella, N.~Vandersickel and H.~Verschelde,
  Phys.\ Rev.\ D {\bf 77}, 071501 (2008).

\bibitem{Dudal:2008sp}
  D.~Dudal, J.~A.~Gracey, S.~P.~Sorella, N.~Vandersickel and H.~Verschelde,
  Phys.\ Rev.\ D {\bf 78}, 065047 (2008).


\bibitem{Dudal:2011gd}
  D.~Dudal, S.~P.~Sorella and N.~Vandersickel,
  Phys.\ Rev.\ D {\bf 84}, 065039 (2011).



\bibitem{Cucchieri:2007rg}
  A.~Cucchieri and T.~Mendes,
  Phys.\ Rev.\ Lett.\  {\bf 100}, 241601 (2008)

\bibitem{Bornyakov:2008yx}
  V.~G.~Bornyakov, V.~K.~Mitrjushkin and M.~Muller-Preussker,
  Phys.\ Rev.\ D {\bf 79}, 074504 (2009).

\bibitem{Dudal:2010tf}
  D.~Dudal, O.~Oliveira and N.~Vandersickel,
  Phys.\ Rev.\ D {\bf 81}, 074505 (2010).


\bibitem{Cucchieri:2011ig}
  A.~Cucchieri, D.~Dudal, T.~Mendes and N.~Vandersickel,
  Phys.\ Rev.\ D {\bf 85}, 094513 (2012).

\bibitem{Oliveira:2012eh}
  O.~Oliveira and P.~J.~Silva,
  Phys.\ Rev.\ D {\bf 86}, 114513 (2012).

\bibitem{Dudal:2012zx}
  D.~Dudal, O.~Oliveira and J.~Rodriguez-Quintero,
  Phys.\ Rev.\ D {\bf 86}, 105005 (2012).

\bibitem{Stingl:1994nk}
  M.~Stingl,
  Z.\ Phys.\ A {\bf 353}, 423 (1996).

\bibitem{Baulieu:2009ha}
  L.~Baulieu, D.~Dudal, M.~S.~Guimaraes, M.~Q.~Huber, S.~P.~Sorella, N.~Vandersickel and D.~Zwanziger,
  Phys.\ Rev.\ D {\bf 82},  025021 (2010).

\bibitem{Windisch:2012sz}
  A.~Windisch, M.~Q.~Huber and R.~Alkofer,
  Phys.\ Rev.\ D {\bf 87}, 065005 (2013).

\bibitem{Dudal:2010cd}
  D.~Dudal, M.~S.~Guimaraes and S.~P.~Sorella,
  Phys.\ Rev.\ Lett.\  {\bf 106}, 062003 (2011).

\bibitem{Roberts:1994dr}
  C.~D.~Roberts and A.~G.~Williams,
  Prog.\ Part.\ Nucl.\ Phys.\  {\bf 33}, 477 (1994).

\bibitem{Bhagwat:2002tx}
  M.~Bhagwat, M.~A.~Pichowsky and P.~C.~Tandy,
  Phys.\ Rev.\ D {\bf 67}, 054019 (2003).

\bibitem{Baulieu:2009xr}
  L.~Baulieu, M.~A.~L.~Capri, A.~J.~Gomez, V.~E.~R.~Lemes, R.~F.~Sobreiro and S.~P.~Sorella,
  Eur.\ Phys.\ J.\ C {\bf 66}, 451 (2010).

\bibitem{Dudal:2013vha}
  D.~Dudal, M.~S.~Guimaraes, L.~F.~Palhares and S.~P.~Sorella,
  arXiv:1303.7134 [hep-ph].


\bibitem{Benic:2012ec}
  S.~Benic, D.~Blaschke and M.~Buballa,
  Phys.\ Rev.\ D {\bf 86}, 074002 (2012).

\bibitem{Fukushima:2012qa}
  K.~Fukushima and K.~Kashiwa,
  arXiv:1206.0685 [hep-ph].


\bibitem{Aouane:2011fv}
 R.~Aouane, V.~G.~Bornyakov, E.~M.~Ilgenfritz, V.~K.~Mitrjushkin, M.~Muller-Preussker and A.~Sternbeck,
  Phys.\ Rev.\ D {\bf 85}, 034501 (2012).


\bibitem{Cucchieri:2011di}
  A.~Cucchieri and T.~Mendes,
  PoS FACESQCD {\bf }, 007 (2010).




\bibitem{Maggiore:1993wq}
  N.~Maggiore and M.~Schaden,
  Phys.\ Rev.\ D {\bf 50}, 6616 (1994).


\bibitem{Baulieu:2008fy}
  L.~Baulieu and S.~P.~Sorella,
  Phys.\ Lett.\ B {\bf 671}, 481 (2009).



\bibitem{Sorella:2009vt}
  S.~P.~Sorella,
  Phys.\ Rev.\ D {\bf 80}, 025013 (2009).

\bibitem{Dudal:2010hj}
  D.~Dudal and N.~Vandersickel,
  Phys.\ Lett.\ B {\bf 700}, 369 (2011).


\bibitem{Dudal:2009xh}
  D.~Dudal, S.~P.~Sorella, N.~Vandersickel and H.~Verschelde,
  Phys.\ Rev.\ D {\bf 79}, 121701 (2009).


\bibitem{Capri:2010hb}
  M.~A.~L.~Capri, A.~J.~Gomez, M.~S.~Guimaraes, V.~E.~R.~Lemes, S.~P.~Sorella and D.~G.~Tedesco,
  Phys.\ Rev.\ D {\bf 82}, 105019 (2010).


\bibitem{Capri:2011wp}
  M.~A.~L.~Capri, A.~J.~Gomez, M.~S.~Guimaraes, V.~E.~R.~Lemes, S.~P.~Sorella and D.~G.~Tedesco,
  Phys.\ Rev.\ D {\bf 83}, 105001 (2011).




\bibitem{vonSmekal:2007ns}
  L.~von Smekal, D.~Mehta, A.~Sternbeck and A.~G.~Williams,
  PoS LAT {\bf 2007}, 382 (2007).

\bibitem{vonSmekal:2008en}
  L.~von Smekal, M.~Ghiotti and A.~G.~Williams,
  Phys.\ Rev.\ D {\bf 78}, 085016 (2008).



\bibitem{Serreau:2012cg}
  J.~Serreau and M.~Tissier,
  Phys.\ Lett.\ B {\bf 712}, 97 (2012).


\bibitem{Lavrov:2011wb}
  P.~Lavrov, O.~Lechtenfeld and A.~Reshetnyak,
  JHEP {\bf 1110}, 043 (2011).

\bibitem{Piguet:1995er}
  O.~Piguet and S.~P.~Sorella,
  Lect.\ Notes Phys.\ M {\bf 28}, 1 (1995).


\bibitem{Barnich:2000zw}
  G.~Barnich, F.~Brandt and M.~Henneaux,
  Phys.\ Rept.\  {\bf 338}, 439 (2000).

\bibitem{Blasi:1990xz}
  A.~Blasi, O.~Piguet and S.~P.~Sorella,
  Nucl.\ Phys.\ B {\bf 356}, 154 (1991).


\bibitem{Dudal:2008rm}
  D.~Dudal, J.~A.~Gracey, S.~P.~Sorella, N.~Vandersickel and H.~Verschelde,
  Phys.\ Rev.\ D {\bf 78}, 125012 (2008).


\end{thebibliography}
\end{document}